\documentclass[journal]{IEEEtran}

\usepackage{subfigure}
\usepackage{graphicx}
\usepackage{dcolumn}
\usepackage{bm}
\usepackage{subfigure}
\usepackage{multicol}
\usepackage{graphicx,epsfig,epstopdf}
\usepackage{cite}
\usepackage{color}
\usepackage{amsmath}
\usepackage{subfigure}
\usepackage{graphicx}
\usepackage{dcolumn}
\usepackage{bm}
\usepackage{subfigure}
\usepackage{graphicx,epsfig,epstopdf}
\usepackage{cite}
\usepackage{resizegather}
\usepackage{soul}
\usepackage{kantlipsum,widetext}
\begin{document}

\bstctlcite{IEEEexample:BSTcontrol}
\title{Analytical Design for Full-space Spatial Power Dividers Using Metagratings}
\author{Hamid~Rajabalipanah, Ali~Abdolali,~\IEEEmembership{Senior Member,~IEEE}
}

\maketitle

\begin{abstract}

We present a rigorous theoretical framework for designing full-space spatial power dividers using metagratings. In our study, the current restrictions of spatial power dividing platforms such as reflection-only performance, operating at normal incidence, and small reflection/refraction angles have been totally relaxed. A modal expansion analysis based on Floquet-Bloch (FB) theorem is established so that a discrete set of spatial harmonics is considered in both reflection and transmission sides of a compound metallic grating in which the unknown coefficients are calculated by applying proper boundary conditions. By eliminating the unwanted scattering harmonics, the proposed metagrating has the ability to realize different functionalities from perfect anomalous refraction to reflection-transmission spatial power dividing, without resorting to full-wave numerical optimizations. The numerical simulations confirm well the theoretical predictions. Our findings not only offer possibilities to realize arbitrary spatial power dividers but also reveal a simple alternative for beamforming array antennas. 
\end{abstract}


\IEEEpeerreviewmaketitle

\section{Introduction}

\IEEEPARstart{T}{he} quest for manipulating the trajectories of electromagnetic (EM) waves has led various researchers to revisit the traditional methodologies in optics community. For several years, the particular problem of coupling an incoming beam into a certain reflection or transmission mode has been approached by means of local phase gradient designs \cite{Yu}. In these structures, the necessity to fine discretization of phase profiles causes high fabrication complexities; Also, the wave manipulation efficiency is limited due to the impedance mismatch between the incident and reflected/refracted waves \cite{Radi}. Particularly, these drawbacks become more highlighted in multibeam applications such as spatial power dividers in which the independent control over the power level of each channel is of great importance \cite{Rajabalipanah2019}. In fact, the performance of current spatial power dividing platforms is accompanied with unwanted parasitic beams and restricted to reflective scenarios, normal incidences, and small deflection angles. 
  
The concept of metagrating has been recently reported by Alu \textit{et. al} \cite{Radi}, based on which, a sparse array of individual particles is believed to realize perfect anomalous reflection and refraction with unitary efficiency. The fundamental physical principle is dictated by the Floquet-Bloch (FB) theorem where, a periodic array of polarizable scatterers diffracts an incident plane wave to a discrete set of propagating and evanescent harmonics. Although many efforts have been paid to the development of metagrating concept, most of them dealt with reflection- or transmission-only configurations, addressed single-beam specifications, and resorted to brute-force full-wave optimizations for meta-atoms \cite{rahmanzadeh, sahar, Rabin, Popov, Epstein}. More specifically, there is no reference addressing reflection-transmission spatial power dividers using analytically-designed metagratings.

In this letter, we present an end-to-end analytical scheme based on FB theorem to design  reflection-transmission spatial power dividers by means of compound metallic gratings. The structure is engineered so that it can realized both symmetric and asymmetric scattering functionalities. To verify the versatility of this systematic approach, four illustrative examples on perfect anomalous refraction and spatial power dividing have been presented, demonstrating that the numerical simulations are in an excellent agreement with our theoretical predictions.

\begin{figure}[t!]
	\centering
	\includegraphics[width=\linewidth]{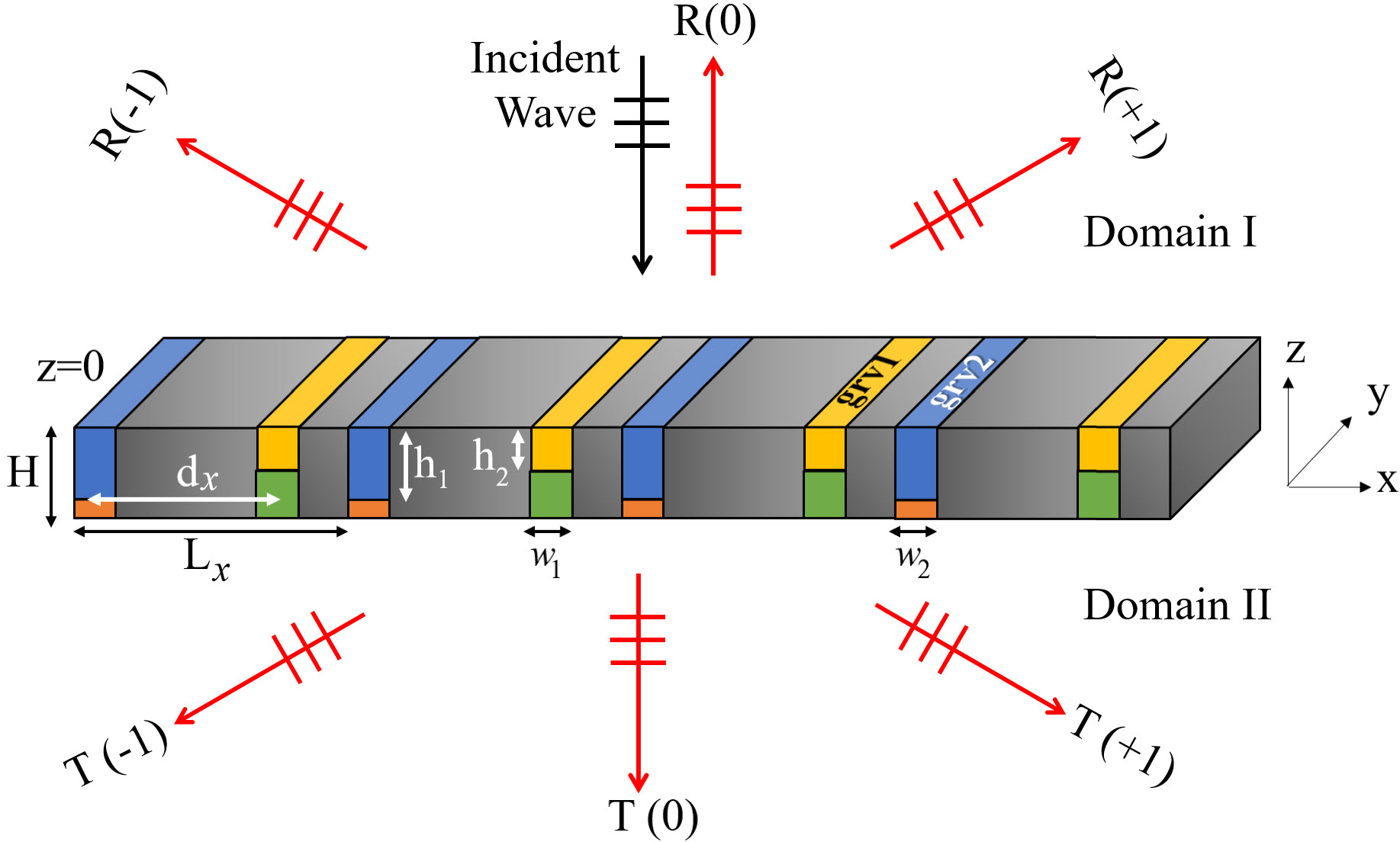}
	\caption{Schematic illustration of full-space spatial power dividing. Each period of the designed metagrating consists of two different grooves filled by cascaded dielectric layers. }
	\label{fig:1}
\end{figure}

\section{Formulation}
Without loss of generality, a 2D configuration ($\partial $/$\partial $$y$=0) is studied comprising periodic pairs of dissimilar grooves with the lattice constant $L_x$. As shown in \textcolor{blue}{Fig. 1}, the metallic grooves (scatterers) are specified with the widths $w_1$, $w_2$ and height $H$ while being partially filled with the dielectric layers $\epsilon_{r1}^{\text{up}}$, $\epsilon_{r2}^{\text{up}}$, $\epsilon_{r1}^{\text{down}}$,$\epsilon_{r2}^{\text{down}}$ of the length $h_1$, $H-h_1$, $h_2$, and $H-h_2$, respectively. The whole structure is surrounded by a medium with dielectric constatnt $\epsilon_0$. The compound metallic gratings are excited by oblique a transverse magnetic (TM)
polarized plane waves propagating along $-z$ direction with $\theta_i$ determining the incident wave angle. According to the FB theorem, the proposed metagrating makes multiple diffractive channels in both transmission and reflection sides, for each of which, the reflection and refraction angles solely depend on the lattice constant while the diffraction efficiency is related to the properties of contributing scatterers. The total fields in the reflection and transmission sides can be written as: 
\begin{align}
&E_{tot,I}^{x}=-{{Y}_{0,I}}{{e}^{j{{k}_{z,0}}z}}{{e}^{-j{{k}_{x,0}}x}}+\\ \nonumber
&~~~~~~~~~~~~~~~~~~~~~~~~\sum\limits_{m}{{{Y}_{m,I}}{{R}_{m}}{{e}^{-j{{k}_{z,m}}z}}{{e}^{-j{{k}_{x,m}}x}}} \\ 
&H_{tot,I}^{y}={{e}^{j{{k}_{z,0}}z}}{{e}^{-j{{k}_{x,0}}x}}+\\ \nonumber &~~~~~~~~~~~~~~~~~~~~~~~~\sum\limits_{m}{{{R}_{m}}{{e}^{-j{{k}_{z,m}}z}}{{e}^{-j{{k}_{x,m}}x}}} \\
&E_{tot,II}^{x}=-\sum\limits_{m}{{{Y}_{m,I}}{{T}_{m}}{{e}^{j{{k}_{z,m}}\left( z+H \right)}}{{e}^{-j{{k}_{x,m}}x}}} \\ 
&H_{tot,II}^{y}=\sum\limits_{m}{{{T}_{m}}{{e}^{j{{k}_{z,m}}\left( z+H \right)}}{{e}^{-j{{k}_{x,m}}x}}} 
\end{align}

in which, ${{Y}_{m}}$$=$${{{k}_{z,m}}}/{\omega {{\varepsilon }_{0}}}$, ${{k}_{z,m}}$$=$$\sqrt{k_{0}^{2}-k_{x,m}^{2}}$, and ${{k}_{x,m}}$$=$${{k}_{0}\sin\theta}+{2m\pi }/{{{L}_{x}}}$ denote the wave admittance and the wavenumbers along z and x directions for $m{\text{th}}$ mode, respectively. Time harmonic dependency of $e^{j\omega t}$ is assumed and suppressed. For operating frequencies less than $f_c=$$c/\max \left[ 2{{w}_{i}}\sqrt{\varepsilon _{r,i}^{up}},2{{w}_{i}}\sqrt{\varepsilon _{r,i}^{down}} \right]\ \left( i=1,\,2 \right)$, each groove can be modelled by a parallel-plate guiding system carrying single TEM propagating mode along forward and backward directions, \textit{i.e.,} 
\begin{align}
& E_{tot,grv1,up}^{x}=-{{Y}_{grv1,up}}{{A}^{+}}{{e}^{j{{k}_{z,grv1,up}}\left( z+{{h}_{1}} \right)}}+\\ \nonumber
&~~~~~~~~~~~~~~~~~~~~~~~~~~~~~~~~{{Y}_{grv1,up}}{{A}^{-}}{{e}^{-j{{k}_{z,grv1,up}}\left( z+{{h}_{1}} \right)}} \\ 
& H_{tot,grv1,up}^{y}={{A}^{+}}{{e}^{j{{k}_{z,grv1,up}}\left( z+{{h}_{1}} \right)}}+{{A}^{-}}{{e}^{-j{{k}_{z,grv1,up}}\left( z+{{h}_{1}} \right)}}\\
&E_{tot,grv1,down}^{x}=-{{Y}_{grv1,down}}{{B}^{+}}{{e}^{j{{k}_{z,grv1,down}}\left( z+{{h}_{1}} \right)}}+\\ \nonumber
&~~~~~~~~~~~~~~~~~~~~~~~~~~~~~~~~{{Y}_{grv1,down}}{{B}^{-}}{{e}^{-j{{k}_{z,grv1,down}}\left( z+{{h}_{1}} \right)}} \\ 
&H_{tot,grv1,down}^{y}={{B}^{+}}{{e}^{j{{k}_{z,grv1,down}}\left( z+{{h}_{1}} \right)}}+{{B}^{-}}{{e}^{-j{{k}_{z,grv1,down}}\left( z+{{h}_{1}} \right)}} 
\end{align} 
for the first groove, and 
\begin{align}
&E_{tot,grv2,up}^{x}=-{{Y}_{grv2,up}}{{C}^{+}}{{e}^{j{{k}_{z,grv2,up}}\left( z+{{h}_{2}} \right)}}+\\ \nonumber
&~~~~~~~~~~~~~~~~~~~~~~~~~~~~~~~~{{Y}_{grv2,up}}{{C}^{-}}{{e}^{-j{{k}_{z,grv2,up}}\left( z+{{h}_{2}} \right)}} \\ 
&H_{tot,grv2,up}^{y}={{C}^{+}}{{e}^{j{{k}_{z,grv2,up}}\left( z+{{h}_{2}} \right)}}+{{C}^{-}}{{e}^{-j{{k}_{z,grv2,up}}\left( z+{{h}_{2}} \right)}} \\ 
&E_{tot,grv2,down}^{x}=-{{Y}_{grv2,down}}{{D}^{+}}{{e}^{j{{k}_{z,grv2,down}}\left( z+h{{}_{2}} \right)}}+\\ \nonumber
&~~~~~~~~~~~~~~~~~~~~~~~~~~~~~~~~{{Y}_{grv2,down}}{{D}^{-}}{{e}^{-j{{k}_{z,grv2,down}}\left( z+h{{}_{2}} \right)}} \\ 
&H_{tot,grv2,down}^{y}={{D}^{+}}{{e}^{j{{k}_{z,grv2,down}}\left( z+h{{ }_{2}} \right)}}+{{D}^{-}}{{e}^{-j{{k}_{z,grv2,down}}\left( z+h{{}_{2}} \right)}} 
\end{align} 
for the second groove. Here, $+$, $-$ superscripts indicate the forward and backward waves propagating inside the grooves, respectively, and ${{k}_{z,grv\,i,up\left( down \right)}}$$=$${{k}_{0}}\sqrt{\varepsilon _{r}^{up\left( down \right)}}$ and $Y_{grvi,up\left( down \right)}^{{}}={{k}_{z,grvi,up\left( down \right)}}/\omega {{\varepsilon }_{0}}\varepsilon _{r}^{up\left( down \right)}$. To find the unknown coefficients, we should impose the configuration's boundary conditions: continuity of the tangential electric and magnetic fields over the groove aperture at $z=0,-H$, $-h_1$ (first groove), and $-h_2$ (second groove) and vanishing the tangential electric fields on any PEC interface at $z=0$ and $z=-H$. After some algebraic manipulations and using the orthogonality of exponential terms in each period, we have: 

\begin{gather}
\begin{bmatrix}    {{G}_{11}} & {{G}_{12}} & {{G}_{13}} & {{G}_{14}}  \\
{{G}_{21}} & {{G}_{22}} & {{G}_{23}} & {{G}_{34}}  \\
{{G}_{31}} & {{G}_{32}} & {{G}_{34}} & {{G}_{34}}  \\
{{G}_{41}} & {{G}_{42}} & {{G}_{43}} & {{G}_{44}}  \\ \end{bmatrix}
\begin{bmatrix} &{{B}^{+}}~~ \\ 
&{{B}^{-}}~~ \\ 
&{{D}^{+}}~~ \\ 
&{{D}^{-}}~~ \\ 
\end{bmatrix}
=
\begin{bmatrix}
&-2M_{0,grv1}^{-}~~ \\ 
& -2M_{0,grv1}^{-} ~~\\ 
& 0 ~~\\ 
& 0 ~~\\ 
\end{bmatrix}
\end{gather}

wherein, 

\begin{widetext}
	\begin{align}
	&G_{1_{2}^{1}}=\sum{\left( \frac{1}{2}\frac{{{Y}_{grv1,up}}}{{{Y}_{m,I}}}M_{m,grv1}^{+}\xi _{1}^{-}\mp\frac{1}{2}\frac{{{Y}_{grv1,down}}}{{{Y}_{m,I}}}M_{m,grv1}^{+}\xi _{1}^{+} \right)M_{m,grv1}^{-}-\frac{1}{2}\frac{{{w}_{grv1}}}{L}\xi _{1}^{+}\pm\frac{1}{2}\frac{{{w}_{grv1}}}{L}\frac{{{Y}_{grv1,down}}}{{{Y}_{grv1,up}}}\xi _{1}^{-}}\\
	&G_{1_{4}^{3}}=\sum{\left( \frac{1}{2}\frac{{{Y}_{grv2,up}}}{{{Y}_{m,I}}}M_{m,grv2}^{+}\xi _{2}^{-}\mp\frac{1}{2}\frac{{{Y}_{grv2,down}}}{{{Y}_{m,I}}}M_{m,grv2}^{+}\xi _{2}^{+} \right)M_{m,grv1}^{-}}\\
	&G_{2_{2}^{1}}=\sum{\left( \frac{1}{2}\frac{{{Y}_{grv1,up}}}{{{Y}_{m,I}}}M_{m,grv1}^{+}\xi _{1}^{-}\mp\frac{1}{2}\frac{{{Y}_{grv1,down}}}{{{Y}_{m,I}}}M_{m,grv1}^{+}\xi _{1}^{+} \right)M_{m,grv2}^{-}}\\
	&G_{2_{4}^{3}}=\sum{\left( \frac{1}{2}\frac{{{Y}_{grv2,up}}}{{{Y}_{m,I}}}M_{m,grv2}^{+}\xi _{2}^{-}\mp\frac{1}{2}\frac{{{Y}_{grv2,down}}}{{{Y}_{m,I}}}M_{m,grv2}^{+}\xi _{2}^{+} \right)M_{m,grv2}^{-}-\frac{1}{2}\frac{{{w}_{grv2}}}{L}\xi _{2}^{+}\pm\frac{1}{2}\frac{{{w}_{grv2}}}{L}\frac{{{Y}_{grv2,down}}}{{{Y}_{grv2,up}}}\xi _{2}^{-}}\\
	&G_{3_{2}^{1}}=\sum{\left( \frac{\pm{{Y}_{grv1,down}}}{{{Y}_{m,I}}}M_{m,grv1}^{+}{{e}^{\pm j{{k}_{z,grv1,down}}\left( -H+{{h}_{1}} \right)}} \right)M_{m,grv1}^{-}-\frac{{{w}_{grv1}}}{L}{{e}^{\pm j{{k}_{z,grv1,down}}\left( -H+{{h}_{1}} \right)}}}\\
	&G_{3_{4}^{3}}=\sum{\left( \,\,\frac{\pm{{Y}_{grv2,down}}}{{{Y}_{m,I}}}M_{m,grv2}^{+}{{e}^{\pm j{{k}_{z,grv2,down}}\left( -H+{{h}_{2}} \right)}}\,\, \right)M_{m,grv1}^{-}}\\
	&G_{4_{2}^{1}}=\sum{\left( \frac{\pm{{Y}_{grv1,down}}}{{{Y}_{m,I}}}M_{m,grv1}^{+}{{e}^{\pm j{{k}_{z,grv1,down}}\left( -H+{{h}_{1}} \right)}} \right)M_{m,grv2}^{-}}\\
	&G_{4_{4}^{3}}=\sum{\left( \,\,\frac{\pm{{Y}_{grv2,down}}}{{{Y}_{m,I}}}M_{m,grv2}^{+}{{e}^{\pm j{{k}_{z,grv2,down}}\left( -H+{{h}_{2}} \right)}}\,\, \right)M_{m,grv2}^{-}-\frac{{{w}_{grv2}}}{L}{{e}^{\pm j{{k}_{z,grv2,down}}\left( -H+h{{ }_{2}} \right)}}}
	\end{align}
\end{widetext}

Here, the up/bottom signs correspond to the up/bottom indices in the left-hand side of equation, and
\begin{align}
\xi _{i}^{\pm }=\pm {{e}^{j{{k}_{z,grvi,up}}{{h}_{i}}}}+{{e}^{-j{{k}_{z,grvi,up}}{{h}_{i}}}}\\
M_{m,grv i}^{\pm }=\frac{1}{L_x}\int\limits_{x\in grv i}^{{}}{{{e}^{\pm j{{k}_{x,m}}x}}dx}
\end{align}

After solving \textcolor{blue}{Eq. (13)}, the transmission and reflection coefficients corresponding to each spatial harmonic can be obtained as: 

\begin{align}
& {{R}_{0}}=1+\frac{1}{2}\frac{{{Y}_{grv1,up}}}{{{Y}_{0,I}}}M_{0,grv1}^{+}{{B}^{+}}\xi_{1}^{-}-\frac{1}{2}\frac{{{Y}_{grv1,down}}}{{{Y}_{0,I}}}M_{0,grv1}^{+}{{B}^{+}}\xi _{1}^{+} \\ \nonumber
& +\frac{1}{2}\frac{{{Y}_{grv1,up}}}{{{Y}_{0,I}}}M_{0,grv1}^{+}{{B}^{-}}\xi _{1}^{-}+\frac{1}{2}\frac{{{Y}_{grv1,down}}}{{{Y}_{0,I}}}M_{0,grv1}^{+}{{B}^{-}}\xi _{1}^{+} \\ \nonumber
& +\frac{1}{2}\frac{{{Y}_{grv2,up}}}{{{Y}_{0,I}}}M_{0,grv2}^{+}{{D}^{+}}\xi _{2}^{-}-\frac{1}{2}\frac{{{Y}_{grv2,down}}}{{{Y}_{0,I}}}M_{0,grv2}^{+}{{D}^{+}}\xi _{2}^{+} \\ \nonumber
& +\frac{1}{2}\frac{{{Y}_{grv2,up}}}{{{Y}_{0,I}}}M_{0,grv2}^{+}{{D}^{-}}\xi _{2}^{-}+\frac{1}{2}\frac{{{Y}_{grv2,down}}}{{{Y}_{0,I}}}M_{0,grv12}^{+}{{D}^{-}}\xi _{2}^{+} \\ \nonumber
\end{align}

\begin{align}
& {{R}_{m\ne0}}=\frac{1}{2}\frac{{{Y}_{grv1,up}}}{{{Y}_{m,I}}}M_{m,grv1}^{+}{{B}^{+}}\xi _{1}^{-}-\frac{1}{2}\frac{{{Y}_{grv1,down}}}{{{Y}_{m,I}}}M_{m,grv1}^{+}{{B}^{+}}\xi _{1}^{+} \\ \nonumber
& +\frac{1}{2}\frac{{{Y}_{grv1,up}}}{{{Y}_{m,I}}}M_{m,grv1}^{+}{{B}^{-}}\xi _{1}^{-}+\frac{1}{2}\frac{{{Y}_{grv1,down}}}{{{Y}_{m,I}}}M_{m,grv1}^{+}{{B}^{-}}\xi _{1}^{+} \\ \nonumber
& +\frac{1}{2}\frac{{{Y}_{grv2,up}}}{{{Y}_{m,I}}}M_{m,grv2}^{+}{{D}^{+}}\xi _{2}^{-}-\frac{1}{2}\frac{{{Y}_{grv2,down}}}{{{Y}_{m,I}}}M_{m,grv2}^{+}{{D}^{+}}\xi _{2}^{+} \\ \nonumber
& +\frac{1}{2}\frac{{{Y}_{grv2,up}}}{{{Y}_{m,I}}}M_{m,grv2}^{+}{{D}^{-}}\xi _{2}^{-}+\frac{1}{2}\frac{{{Y}_{grv2,down}}}{{{Y}_{m,I}}}M_{m,grv12}^{+}{{D}^{-}}\xi _{2}^{+} \\ \nonumber
\end{align}
\begin{align}
& {{T}_{m}}=\frac{{{Y}_{grv1,down}}}{{{Y}_{m,I}}}{{B}^{+}}M_{m,grv1}^{+}{{e}^{j{{k}_{z,grv1,down}}\left( -H+{{h}_{1}} \right)}}\\ \nonumber 
&~~~~~-\frac{{{Y}_{grv1,down}}}{{{Y}_{m,I}}}{{B}^{-}}M_{m,grv1}^{+}{{e}^{-j{{k}_{z,grv1,down}}\left( -H+{{h}_{1}} \right)}} \\ \nonumber
& ~~~~~+\frac{{{Y}_{grv2,down}}}{{{Y}_{m,I}}}{{D}^{+}}M_{m,grv2}^{+}{{e}^{j{{k}_{z,grv2,down}}\left( -H+{{h}_{2}} \right)}}\\ \nonumber
&~~~~~-\frac{{{Y}_{grv2,down}}}{{{Y}_{m,I}}}{{D}^{-}}M_{m,grv2}^{+}{{e}^{-j{{k}_{z,grv2,down}}\left( -H+{{h}_{2}} \right)}} \\ \nonumber
\end{align} 

Finally, the diffraction efficiencies can be calculated through:
\begin{align}
&\eta_{R,m}={{R}_{m}}R_{m}^{*}\left( \frac{{{k}_{z,m}}}{{{k}_{z,0}}} \right)\\ 
&\eta_{T,m}={{T}_{m}}T_{m}^{*}\left( \frac{{{k}_{z,m}}}{{{k}_{z,0}}} \right)
\end{align} 

The presented analysis can be simply extended to metagratings consisting of more than two grooves in each period. For the sake of briefness, the details are not given here. 

\section{Results and Discussion}
In accordance with the FB theorem, the reflected and transmitted waves consist of a superposition of spatial harmonics departing at angles 
\begin{align}
\sin \theta _{m,R(T)}-\sin {{\theta }_{i}}=\frac{2m\pi }{{{k}_{0}}{{L}_{x}}}  ~~~~~m=0,\pm1, \pm2, ...
\end{align} 

Any power coupling between the spatial harmonics must be allowed by the FB theorem, \textit{i.e.,} the diffraction modes of interest should not be evanescent so that they can contribute to the far-field scattering. For instance, let us consider a metagrating with the structural parameters $\epsilon_{r1}^\text{up}=2$, $\epsilon_{r1}^\text{down}=1$, $\epsilon_{r2}^\text{up}=1, \epsilon_{r2}^\text{down}=2$, $w_1=0.12L_x$, $w_2=0.2L_x$, $h_1=0.3L$, $h_2=0.6L_x$, $d_x=0.4L_x$, $H=1.08L_x$, and $L_x$=$\lambda/\sin(|\theta_{\pm1}|)=\lambda/\sin(60^\circ)$ ($f$=13 GHz) making only three $m=0, \pm1$ spatial harmonics propagative. The desired inhomogeneity along inside the grooves can be simply realized through drilling holes in a homogeneous host dielectric \cite{Rajab2018}. A commercial full-wave software, CST Microwave Studio is utilized to characterize the metagrating channels in which the unit cell boundary conditions are applied to x and y directions while Floquet ports are assigned the walls along z direction. The structure is illuminated by TM-polarized plane waves along $\theta_i=0^\circ$ direction. The analytical and numerical diffraction efficiencies for each of three spatial harmonics in the reflection and transmission sides are plotted in \textcolor{blue}{Fig. 2}. The center frequency is considered as $f$=10 GHz for normalization purpose. The excellent agreement between the calculated and simulated results confirms that the proposed analytical scheme can be employed for optimizing the structure to achieve different functionalities. \\
\begin{figure}[b!]
	\centering
	\includegraphics[width=60mm]{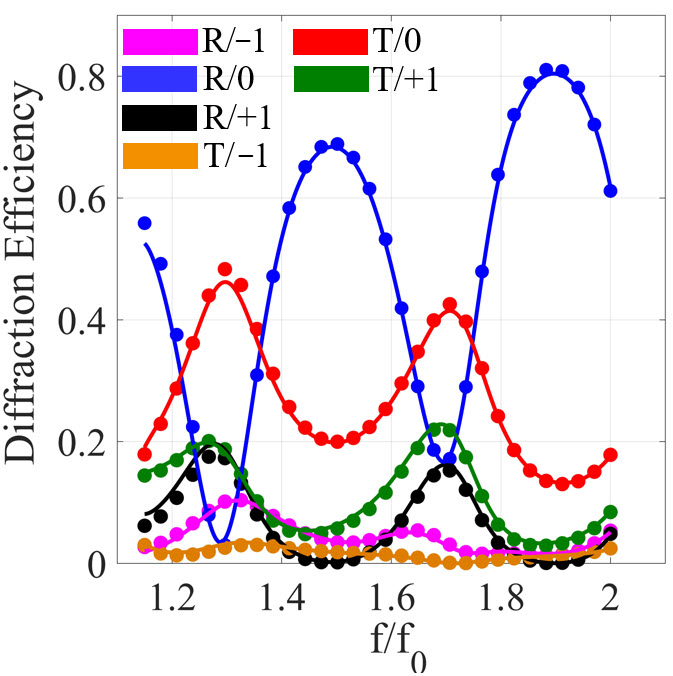}
	\caption{The diffraction efficiency spectra of different propagating space harmonics in the reflection and transmission sides of the designed metagrating.  }
	\label{fig:2}
\end{figure}

 \begin{figure*}[h!]
 	\centering
 	\includegraphics[width=190mm]{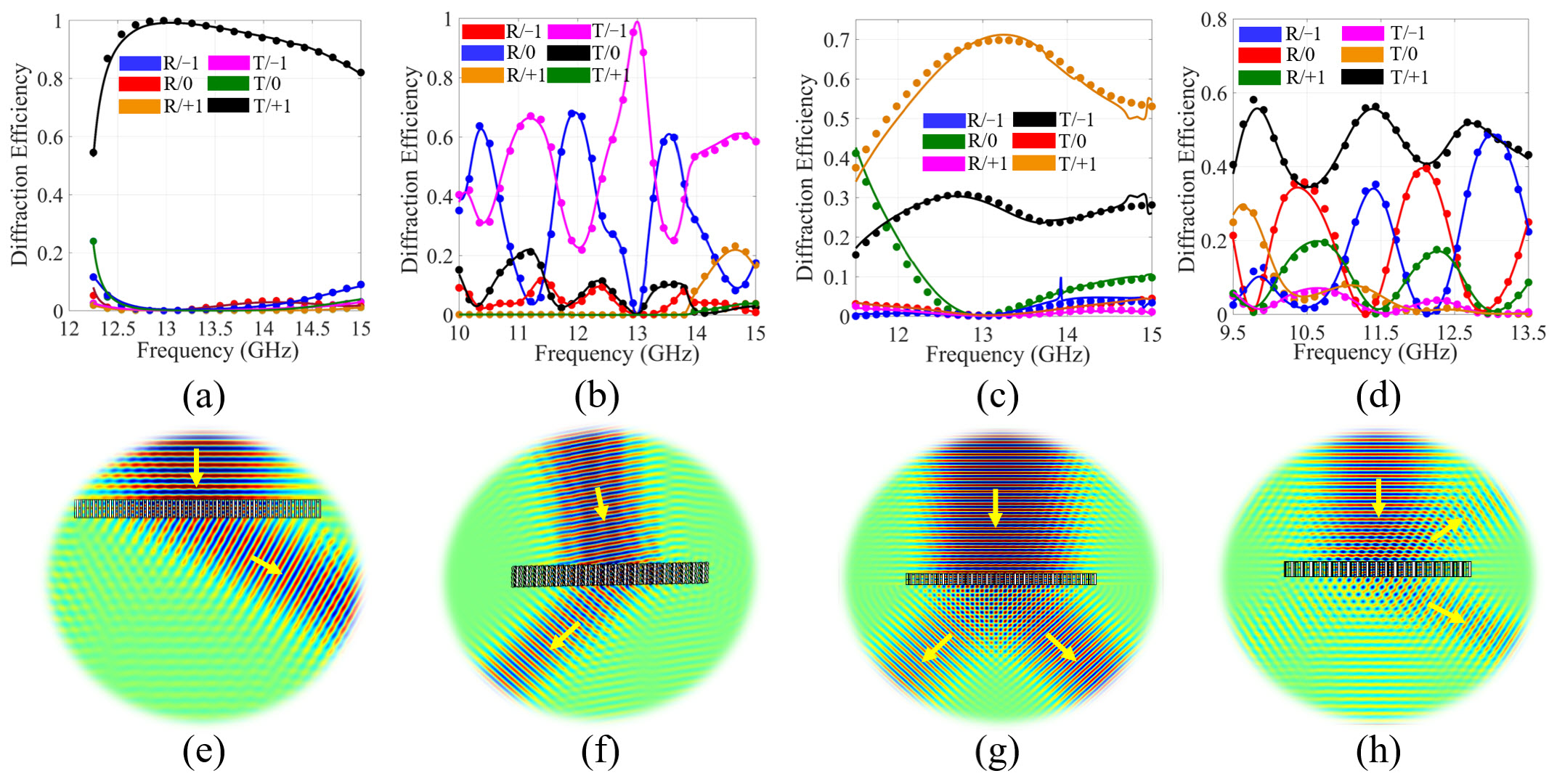}
 	\caption{Diffraction efficiency spectra and the electric field distributions related to the problem of: (a), (e) perfect anomalous transmission under normal illumination achieved by the optimum parameters $\epsilon_{r1}^\text{up}=1$, $\epsilon_{r1}^\text{down}=1.7$, $\epsilon_{r2}^\text{up}=1$, $\epsilon_{r2}^\text{down}=3$, $\epsilon_{r3}^\text{up}=1.2$, $\epsilon_{r3}^\text{down}=2.8$,, $w_1=0.2L_x$, $w_2=0.2L_x$, $w_3=0.2L_x$, $h_1=0.69L_x$, $h_2=0.66L_x$, $h_3=0.18L_x$, $d_{x1}=0.34L_x$, $d_{x2}=0.33L_x$, $H=1.41L_x$, and $L_x$=$\lambda/\sin(|70^\circ|)$. (b), (f) perfect anomalous transmission under $\theta=10^\circ$ illumination achieved by the optimum parameters $\epsilon_{r1}^\text{up}=2$, $\epsilon_{r1}^\text{down}=2$, $\epsilon_{r2}^\text{up}=1$, $\epsilon_{r2}^\text{down}=2$, $\epsilon_{r3}^\text{up}=2$, $\epsilon_{r3}^\text{down}=1.5$,, $w_1=0.2L_x$, $w_2=0.12L_x$, $w_3=0.16L_x$, $h_1=0.89L_x$, $h_2=0.52L_x$, $h_3=0.17L_x$, $d_{x1}=0.275L_x$, $d_{x2}=0.374L_x$, $H=2.37L_x$, and $L_x$=$\lambda/(\sin(|10^\circ|)+\sin(|45^\circ|))$. (c), (g) spatial power dividing in the transmission side achieved by the optimum parameters $\epsilon_{r1}^\text{up}=1.5$, $\epsilon_{r1}^\text{down}=1.15$, $\epsilon_{r2}^\text{up}=1.8$, $\epsilon_{r2}^\text{down}=1.15$, $\epsilon_{r3}^\text{up}=1$, $\epsilon_{r3}^\text{down}=2.5$,, $w_1=0.23L_x$, $w_2=0.15L_x$, $w_3=0.26L_x$, $h_1=0.66L_x$, $h_2=0.64L_x$, $h_3=0.18L_x$, $d_{x1}=0.21L_x$, $d_{x2}=0.38L_x$, $H=1.27L_x$, and $L_x$=$\lambda/\sin(|60^\circ|)$. (d), (h) spatial power dividing in both transmission and reflection sides achieved by the optimum parameters $\epsilon_{r1}^\text{up}=1$, $\epsilon_{r1}^\text{down}=2.6$, $\epsilon_{r2}^\text{up}=1$, $\epsilon_{r2}^\text{down}=1.8$, $\epsilon_{r3}^\text{up}=2.1$, $\epsilon_{r3}^\text{down}=2.1$,, $w_1=0.12L_x$, $w_2=0.26L_x$, $w_3=0.23L_x$, $h_1=0.19L_x$, $h_2=0.13L_x$, $h_3=0.76L_x$, $d_{x1}=0.276L_x$, $d_{x2}=0.253L_x$, $H=1.83L_x$, and $L_x$=$\lambda/\sin(|45^\circ|)$.}
 	\label{fig:3}
 \end{figure*}

\subsection{Perfect Anomalous Refraction}

In this section, we follow the theoretical scheme described in the previous section to find the best structural parameters, without requiring even a single optimization in a full-wave solver.  Without loss of generality, we assume that all spatial harmonics except $m=0, \pm1$ are evanescent in both reflection and transmission windows. Based on \textcolor{blue}{Eq. (29)}, this assumption is quite logical when $\lambda/(1-\sin\theta_i)<L_x<2\lambda/(1+\sin\theta_i)$. Meanwhile, the proper lattice constant to align the $mth$ channel along $\theta_m$ direction in either reflection or transmission side is $L_x=m\lambda/(\sin\theta_{m}-\sin\theta_i)$. Our objective is to design a metallo-dielectric metagrating that routes a plane wave with a given angle of incidence in to a prescribed transmission channel, without generating spurious diffractions. In order to do so, we seek to properly set the two degrees of freedom available in our configuration, namely, the thickness and permittivity values of dielectric layers, and the width and relative position of grooves. The metagrating is assumed to have three grooves in each period for realizing the desired wave transformations with sufficient degrees of freedom. To realize perfect anomalous refraction upon illuminating by a nromal incidence, the goal function of the optimization procedure is to minimize $|\eta_{T,+1}-1|^2$ so that the whole incident power is channeled to the $m=+1$ harmonic. The lattice constant is chosen according to the desired channel orientation as $L_x$=$\lambda/\sin(|70^\circ|)$ ($f$=13 GHz). The diffraction efficiencies are depicted in \textcolor{blue}{Fig. 3a}, illustrating a perfect refraction with the efficiency of \%99.9. The optimum parameters are given in caption of the same figure. To have a better interpretation, COMSOL Multiphysics sofwtare has been used to demonstrate the perfect refraction phenomenon. PML boundaries have been utilized to surround the simulation domain. The electric fields at $f$=13 GHz have been captured and shown in \textcolor{blue}{Fig. 3e} which clearly remark that the normal incident wave is efficiently refracted into the $m=+1$ spatial harmonic with the angle $\theta_{+1}=70^\circ$. We repeated the design process for another case in which an oblique plane wave $\theta$=10$^\circ$ excites the metagrating and is desired to be perfectly transmitted toward $\theta$=$-$45$^\circ$ direction ($m=-1$ mode). \textcolor{blue}{Figs. 3b, f} display the diffraction efficiency and the electric fields pertaining to such wave transformation. As can be deduced, the perfect anomalous refraction has been successfully realized with the efficiency of \%99.8.    

\subsection{Spatial Power Dividers}
Though being essential in various beamforming applications, a few studies have contributed to address configurations producing multiple beams with adjustable power ratio levels. However, the previous proposals are based on local designs and have been restricted to reflection scenarios and normal illuminations, require local amplitude modulation, possess limited efficiencies especially at near grazing angles due to the phase gradient nature, and also, lacks front-to-end analytical designs \cite{Rajabalipanah2019, tie jun}. Here, we will demonstrate that how our analytically designed metagrating serve to accomplish arbitrary spatial power dividing between different transmission and reflection channels. Given a normal illumination, we seek for the best group of parameters whereby \%30 and 
\%70 of the input power are transformed into the $m=-1$ and $m=+1$ transmission modes, respectively, with $|\theta_{\pm1}|=60^\circ$. In fact, the fitness function is defined as $|\eta_{T,-1}-0.3|^2+|\eta_{T,+1}-0.7|^2$. The corresponding diffraction efficiency spectra are illustrated in \textcolor{blue}{Fig. 3c}. Also, the optimum parameters are given in the caption of same figure. The results demonstrate that, in the vicinity of the operating frequency $f$=13 GHz, the spatial power divider successfully allocates \%29.2 and \%70.5 of the incident power to $m=-1$ and $m=+1$ harmonics in the reflection and transmission sides, respectively. Moreover, further illustration of this spatial dividing can be observed in \textcolor{blue}{Fig. 3g}, where the electric fields clearly show the routes of scattered waves. As another  spatial power dividing example, the metagrating excited normally is optimized to re-direct \%50 and \%50 of the incident power to $m=+1$ modes of the reflection and transmission sides, respectively. The orientation of channels are set toward 45$\circ$ direction. The error function is defined as $|\eta_{T,+1}-0.5|^2+|\eta_{R,+1}-0.5|^2$. \textcolor{blue}{Figs. 3d, h} indicate the diffraction efficiencies and the electric field distribution corresponding to this specific wave transformation, respectively, both of which, corroborate well the performance of the designed metagrating at $f$=13 GHz.

\section{Conclusion}
In summary, this letter was aimed at presenting an end-to-end analytical scheme for designing full-space spatial power dividers through engineering the diffraction modes of a compound metallic grating. In our study, the restrictions of the previous proposals such as reflection-only performance, operating upon normal illuminations, and small beam deflection angles have been relaxed. In an excellent accordance with the numerical results, our theoretical study illustrates that the designed metagrating can serve to realize different functionalities from perfect anomalous refraction to full-space spatial power dividing, without even one 3D full-wave optimization.

\end{document}